# Game Development Software Engineering Process Life Cycle: A Systematic Review

*Saiqa Aleem*[a,*], *Luiz Fernando Capretz*[a], *Faheem Ahmed*[b]

[a] *Department of Electrical & Computer Engineering, University of Western Ontario, London, Ontario, Canada N6A 5B9*
[b] *Department of Computing Science, Thompson Rivers University, Kamloops, British Columbia, Canada, V2C 0C8*

**Abstract**

Software game is a kind of application that is used not only for entertainment, but also for serious purposes that can be applicable to different domains such as education, business, and health care. Although the game development process differs from the traditional software development process because it involves interdisciplinary activities. Software engineering techniques are still important for game development because they can help the developer to achieve maintainability, flexibility, lower effort and cost, and better design. The purpose of this study is to assesses the state of the art research on the game development software engineering process and highlight areas that need further consideration by researchers. In the study, we used a systematic literature review methodology based on well-known digital libraries. The largest number of studies have been reported in the production phase of the game development software engineering process life cycle, followed by the pre-production phase. By contrast, the post-production phase has received much less research activity than the pre-production and production phases. The results of this study suggest that the game development software engineering process has many aspects that need further attention from researchers; that especially includes the postproduction phase.

**Keywords:** Software Game; Video game; Online game; Systematic review; Development software engineerong proces.

* Corresponding author.
E-mail addresses: saleem4@uwo.ca
Phone number: +1-226-977-0417; Fax number: +1-519-850-2436

# 1. Introduction

With the rapid advancement of computer technology, the significance of software engineering in our daily lives is increasing. It affects every aspect of our lives today, including working, living, learning, and education. A new and popular mode of entertainment and an important application of technology are software games, which have become increasingly accepted by people of all ages. In today's culture, technology is easily accessible and has become more convenient; more and more people like to play games and are also becoming motivated to design their own games. Katie Salen and Eric Zimmerman (2003) defined "game is a software application in which one or more players make decisions by controlling game objects and resources, in the pursuit of its goal". Software games are software applications that are installed on hardware devices such as video game consoles, computers, handheld devices, and Personal Digital Assistants (PDAs). Software games have now become a worldwide creative industry, but because of the multidisciplinary activities required, their development is a very complex task.

The multidisciplinary nature of the processes that combine sound, art, control systems, artificial intelligence (AI), and human factors, also makes the software game development practice different from traditional software development. However, despite the high complexity of the software engineering development process, the game industry is making billions of dollars in profit and creating many hours of fun (PWC, 2011-2014 outlook). The software game market throughout the world has grown by over 7%–8% annually and has reached sales of around $5.5 billion in 2015 (SUPERDATA, 2015). Newzoo (2015) has also reported that the world-wide digital game market will reach $113.3 billion by 2018.

Creation of any game involves cross-functional teams including designers, software developers, musicians, script writers, and many others. Also, Entertainment Software Association facts (2014; 2015) reports highlighted the latest trends about the software game industry. Therefore, game development careers have currently become highly challenging, dynamic, creative, and profitable (Liming and Vilorio, 2011). The ability to handle complex development tasks and achieve profitability does not happen by chance, but rather a common set of good practices must be adopted to achieve these goals. The game industry can follow the good and proven practices of traditional software engineering, but only a clear understanding of these practices can enhance the complex game development engineering process.

The computer game domain covers a great variety of player modes and genres (Gredler, 1995; Gredler, 2004; Rieber, 2005). The complexity of software games has posed many challenges and issues in software development engineering process because it involves diverse activities in creative arts disciplines (storyboarding, design, refinement of animations, artificial intelligence, video production, scenarios, sounds, marketing, and, finally, sales) in addition to technological and functional requirements (Keith, 2010). This inherent diversity leads to a greatly fragmented domain from the perspectives of both underlying theory and design methodology. The software game literature published in recent years has focused mainly on technical issues. Issues of game production, development, and testing reflect only the general software-engineering state of the art. Pressman (2001) states that a game is a kind of software that entertains its users, but game development software engineering faces many challenges and issues if only a traditional software-development process is followed (Kanode and Haddad, 2009; Petrillo et al., 2009). Some studies have proposed a Game Development Software Engineering (GDSE) process life cycle that provides guidelines for the game development software engineering process (Hendrick, 2014; Blitz game studio, 2014; McGrath, 2014; Chandler, 2010; Ramdan and Widyani, 2013). However, the proposed GDSE process life cycle development phases do not ensure a quality development process.

A GDSE process is different from a traditional software development engineering process, and all phases of the proposed GDSE process life cycle can be combined into three main phases: pre-production, production, and post-production. The pre-production phase includes testing the feasibility of target game scenarios, including requirements engineering marketing strategies; the production phase involves planning, documentation, and game implementation scenarios with sound and graphics. The last phase post-production involves testing, marketing, and game advertising. Because of high competition and extreme market demand, game development companies sometimes reduce their development process so they can be first to market (Kaitilla, 2014). This reduction of the development process definitely affects game quality. Because of these types of complex project-management tasks, the game development software engineering process diverges from traditional software development. Therefore, it becomes important now to investigate the challenges or issues faced by game development organizations in developing good quality games. This systematic literature review is the first step towards identifying the research gaps in the GDSE field.

**1.1 Related work**

Managing GDSE process life cycle has become a much harder process than anyone could have initially imagined, and because of the fragmented domain, no clear picture of its advancement can be found in the literature. A systematic literature review provides a state of the art examination of an area and raises open research questions in a field, thus saving a great deal of time for those starting research in the field. However, to the best of the authors' knowledge, no systematic literature review has been reported for GDSE process life cycle. Many researchers have adopted the systematic literature review approach to explore different aspects in software games. Boyle et al. (2012) conducted a systematic literature review to explore the engagement factor in entertainment games from a player's perspective. In this study, 55 papers were selected to perform the systematic literature review. The study highlighted the different aspects of engagement factors with entertainment games; these include subjective feelings of enjoyment, physiological responses, motives, game usage, player loyalty, and the impact of playing games on a player's life. Connolly et al. (2012) explored 129 papers to report the impacts and outcomes of computer and serious games with respect to engagement and learning by using the systematic literature review approach.

Another study also reported the importance of engagement in digital games by using a systematic literature review approach. Osborne-O'Hagan et al. (2014) performed a systematic literature review on software development processes for games. A total of 404 studies were analyzed from industry and academia and different software development adoption models used for game development were discussed. The findings of the study were that qualitative studies reported more agile practices than the hybrid approach. The quantitative studies used an almost hybrid approach. We also noted that lightweight agile practices such as Scrum, XP, and Kanban – are suitable where innovation and time to market is important. A risk-driven spiral approach is appropriate for large projects. Only one systematic study was performed related to research on software engineering practices in the computer game domain rather than GDSE process life cycle (Ampatzoglou and Stamelos, 2010).

This study mainly review the existing evidence in the literature concerning the GDSE process research and suggest areas for further investigation by identifying possible gaps in current research. Furthermore, the aim of this study is to cover the state of the art for the GDSE process life cycle, and to accomplish this, an evidence-based research paradigm has been used. In the software engineering field, possible use of an evidence-based paradigm have been proposed by Dyba et al. (2005) and Kitchenham et al. (2004). The Systematic Literature Review (SLR) research paradigm constitutes the first step in an evidence-based paradigm research process, and its guidelines for performing systematic research are thoroughly described by Breton et al. (2007) and Kitchenham (2004).

The rest of the paper is organized as follows: Section 2 provides the research background and Section 3 describes the methodology used for the systematic literature review as described by Breton et al. (2007). Section 4 presents the statistics for the primary studies, Section 5 answers various research questions, Section 6 discuss the external threats to validity, and, finally, Section 7 concludes the presentation.

## 2. Research Background

In the software development industry, software games are gaining importance because they are not only used for entertainment, but also for serious purposes that can be applicable to different domains such as education, business, and health care. Along with their applicability to different domains, their revenue has also been increasing. Games software earned three times more revenue than any other software product in 2012 (Nayak, 2013).

Robin (2009) defines a *development method* as a systematized procedure to achieve the goal of producing a working product within budget and on schedule. A number of methodologies used for game development and design (Novak, 2008; Castillo & Novak, 2008). The first is the waterfall method, which is also commonly used in traditional software development. Unlike game projects, once the pre-production phase is completed, production phase activities are performed in a "waterfall" manner. First, the activities are segregated based on functionalities and assets, and then they are assigned to their respective teams. The requirements team spent a significant amount of time in functionality definition and front-end activities, which implies a late implementation of level and mechanisms (Schwaber & Beedle, 2002). However, in the waterfall method, it is difficult to reverse any activity (Flood, 2003).

The second development methodology is the agile method that is commonly used for game development. These methods are highly iterative and not documentation-centric. The production phase is divided into small iterations and focusses on the most crucial features. During the beginning phase of each iteration, the whole team meets and sets clear objectives. At the end of each iteration, results are communicated to clients. These methods support different team cycles and dynamics through daily meetings. The most used agile methodologies in game development are *extreme programming* (XP), rapid prototyping, and Scrum (Godoy & Barbosa, 2010).

The unified development process (Kruchten, 2000) is another traditional SE method, which focusses more on analyzing requirements and converting them into functional software components. The requirement analysis document includes a definition of the game concept, use cases, and assets definitions (Schwaber & Beedle, 2002). The method includes five disciplines: requirements, analysis, design, implementation, and testing. The unified process is based on a philosophy of four key elements: iterative and incremental, use case-driven, architecture-centric, and risk-driven.

Kanode and Haddad (2009) stated that an important, but incorrect, assumption was made that GDSE follows the waterfall method. More recently, researchers have agreed that it must follow the incremental model (Munassar and Govardhan, 2011) because it combines the waterfall method with an iterative process. A major concern, reported by Petrillo et al. (2009), was that very poor development methodologies are commonly used by developers for software creation in the game industry. The GDSE appears as a question in many forms attempting to determine what types of practices are used. However, there is no single answer to this question. Few researchers have explored GDSE practices and then tried to answer questions like the phases of the GDSE process life cycle. Blitz game studios (2014) proposed six phases for the GDSE process life cycle: Pitch (initial design and game concept), Pre-production (game design document), Main production (implementation of game concepts), Alpha (internal testers), Beta (third-party testers), and the Master phase (game launch). Hendrick (2014) proposed a five-phase GDSE process life cycle consisting of Prototype (initial design prototype), Pre-production (design document), Production (asset creation, source code, integration aspects), Beta (user feedback), and, finally, the Live phase (ready to play). McGrath (2014) divided the GDSE process life cycle into six phases: Design (initial design and game design document), Develop/redevelop (game engine development), Evaluate (if not passed, then redevelop), Test (internal testing), Review release (third-party testing), and Release (game launch). Another GDSE process life cycle proposed by Chandler (2010) consisted of four phases: Pre-production (design document and project planning), Production (technical and artistic), Testing (bug fixing), and, finally, the Post-production phase (post-mortem activities). The latest GDSE process life cycle in 2013 proposed by Ramadan and Widyani (2013) was based on the four GDSE process life cycles previously described. They proposed six phases: Initiation (rough concept), Pre-production (creation of game design and prototype), Production (formal details, refinement, implementation), Testing (bug reports, refinement testing, change requests), Beta (third-party testers), and Release (public release).

In traditional software engineering, the development phase usually involves activities such as application design and its implementation; the production phase is when the software actually runs and is ready for use. However, in the GDSE process lifecycle, the production phase includes the development process, which is the pre-production phase of the traditional software engineering process, and the production phase of traditional software engineering is actually the post-production phase of the GDSE process life cycle (Bethke, 2003). Therefore, the GDSE process life cycle is different from the traditional software engineering process, and many researchers have studied the challenges faced by this domain (Kanode and Haddad, 2009). The most prominent observation made in these studies is that to address the challenges faced by the GDSE process life cycle, more rigorous software engineering strategies must be used. Most researchers have explicitly compared the software engineering process with the GDSE process, but none of them has studied complete GDSE process life cycle and research topics under this domain in detail. This study will provide evidence on these topics and their differences from the traditional software engineering process. In this paper, the GDSE process phases were divided into three phases for basic understanding: Preproduction, Production, and Post-production. Efforts were made to classify these further based on studies found in the literature. The primary contribution of this paper is that it is the first SLR that addresses these GDSE process life cycle research topics and highlights the topics that need further attention by researchers.

## 3. Review Methodology

In this work, the conceptual description of the SLR process presented by Kitchenham (2004) was used to investigate the research intensity for each phase of the GDSE process life cycle. Conceptually, SLR provides an opportunity for researchers to collect empirical evidence from the existing literature about a formulated research question. Although most authors followed the general SLR guidelines provided by Kitchenham (2004), there were slight variations in the description and presentation of the conceptual process layout. The generic SLR guidelines stated by Kitchenham (2004) are further elaborated here, and the overall process is described as a set of activities The research process has been adopted for this study described by Kitchenham and Charters (2007). There are mainly three phases of the review and the steps associated with each phase are shown in fig. 1.

**3.1 Planning Phase (Step 1 -4)**

This study started by selecting a topic, at which point the study objectives were also clearly defined and the boundaries of the domain delineated.

### *3.1.1 Selection of Topic and Research Questions*

Selecting a topic for SLR is of crucial importance because many factors such as individual or community interest, research gaps, and research impact contribute to shaping research questions on the topic. Our understanding of the GDSE process life cycle is continuously evolving (Kitchenham et al., 2010), and many areas in this field lack generalized evidence. It is critically important for the game industry to identify a quality-driven GDSE process. Several studies have investigated different phases of the GDSE process life cycle, but they do not offer systematic, comprehensive, and thorough methodological research specific to this topic.

In this review, studies from 2000 to 2015 will be explored to answer the following research questions:

*Research Question (RQ1):* What is the intensity of research activity on the GDSE process life cycle?

*RQ2:* What topics are being researched in the pre-production, production, and post-production phases?

*RQ3:* What research approaches are being used by researchers in the software game domain?

*RQ4:* What empirical research methods are being used in the software game domain?

The number of publications has been identified by the research group to address RQ1. A graphical representation has been used to represent the increase or decrease in the number of publications per year as a measure of research activity. To address RQ2, RQ3, and RQ4, each study selected has been affiliated to a research topic, to a certain approach, and to a specific methodology used for the research. Details of this classification into corresponding categories are discussed in section 3.2.4.

### 3.1.2 Review Team & Protocol Establishment

A multidisciplinary team is needed to perform a high-quality scientific SLR. To enhance the thoroughness and minimize the potential bias of a study, an SLR is normally undertaken by more than one reviewer. The SLR team for this review was made up of three people. Two people were designated as principal reviewers (Second expert report by American institute 2011). One person was also selected as the project leader to handle additional administrative tasks such as team communication, points of contact, meeting arrangements and documentation, task assignment and follow-up, and quality assurance. Table 1 details the tasks required for the SLR process and reviewer's involvment.

Table I. Reviewers' involvement in SLR tasks

| Task | Team members involved |
|---|---|
| Development of review protocol | Principal reviewers |
| Development of search strategy | All |
| Assessment of papers, including relevance and study design | Other reviewer |
| Data extraction | Other reviewer |
| Data analysis | Principal reviewers |
| Final SLR report | All |
| SLR update | All |

In order to ensure the review could be replicated and to reduce researcher bias a review protocol and it's evaluation procedure was developed at step 3 and 4. The final review protocol is discussed in the following sections 3.2.1 to 3.2.4 (Steps 5-9 incl.).

**3.2 Conducting Phase (Step 5-9)**

*3.2.1 Search Strategy*

In the SLR, the search procedure is based on an online search. The search strategy for an SLR is a plan to construct search terms by identifying populations, interventions, and outcomes. Key terms are combined together to created different groups in order to form search strings. Each group comprise of terms that are either different forms of the same word, synonyms, or terms that have similar or related semantic meaning within the domain. Table 2 depicts the followed approach.

In order to retrieve different sets of relevant literature, four groups are designed. The main objective of this grouping is to find the literature that is the intersection of the groups as shown in Fig. 2.

Table 2 Search terms and corresponding groups

| Number of Terms | Group 1 Software Games | Group 2 Development | Group 3 Lifecycle | Group 4 Process |
|---|---|---|---|---|
| **Term 1** | Digital Games | Advancements | Design | Progression |
| **Term 2** | Video Games | Steps | Requirement Engineering | Method |
| **Term 3** | Computer Games | Evolve | Implementation | Model |
| **Term 4** | Online Games | Project | Evaluation | |
| **Term 5** | Serious Games | | Testing | |
| **Term 6** | Educational Games | | Maintenance | |
| **Term 7** | Learning Games | | | |

The search strategy was implemented by applying the "AND" and "OR", where the "OR" operator is used within the Group and the "AND" is used between the groups. According to Table 2, the following search string will capture the structure:

(**Group 1:** [Software game] **OR** [Digital game] **OR** [Video game] **OR** [Computer game] **OR** [Online Game] **OR** [Serious games] **OR** [Educational Games] **OR**

[Learning Games])

**AND**

(**Group 2:** [Development] **OR** [Advancement] **OR** [Steps] **OR** [Evolve] **OR** [Project])

**AND**

(**Group 3:** [Life cycle] **OR** [Design] **OR** [Implementation] **OR** [Requirements Engineering] **OR** [Testing] **OR** [Evaluation] **OR** [Maintenance])

**AND**

(**Group 4***:* [Process] **OR** [Progression] **OR** [Method] **OR** [Model]).

Therefore, "*Software game development lifecycle process*", "*Computer game development design process*" and "*video game testing process*" are some examples of the search strings and similar way different search strings were formed in order to capture all relevant studies."

To ensure that all relevant research concerning this area of study was reviewed, journals and conferences from 2000 to 2015 were covered, using as sources IEEE Explorer, ACM Digital Library, Science Direct Elsevier, Taylor & Francis, Google Scholar, and Wiley Publications. If the information required, as indicated on the form shown in Table 3, was not explicitly present in the potential study, then that paper was peer-reviewed by all team members and, after discussion, validated for correctness. Otherwise, each paper was reviewed by one reviewer. Each study involved some general information and some specific information, as indicated on the form.

Table 3 Publication-specific data

| Specific Information about a Research Article | |
|---|---|
| **A) Research Methodology used in SLR** | |
| Empirical | |
| Descriptive | |
| Exploratory | |
| **B) Empirical research methods** | |
| Experiment | |
| Survey | |
| Case study | |
| **C) Type of publication** | |
| Journal | |
| Conference | |
| Workshop | |
| **D) Research activity per country** | |

| Country name | Number of publications |
|---|---|
| **E) Year of Publication** | |
| 2000–2005 | |
| 2006–2010 | |
| 2010–2015 | |

### *3.2.2 Pilot Selection & Data Extraction*

The research study selection and data extraction was based on the following coverage criteria:

*Inclusion Criteria for Study*: For SLR, articles and research papers from 2000 to 2015 were included, and to evaluate their suitability, the following criteria were analyzed:

- The study should be thoroughly reviewed by at least one of the reviewers.
- Only the following types of studies were considered: case studies, theoretical papers, and empirical analysis surveys.
- The full text of the article should be available.
- If any article identifies any challenges and problems in software games, that article is included as a review.
- Studies that describe motivation for game application.

*Study Exclusion Criteria*: The following criteria were used to determine articles to be excluded:

- Articles published on company Web sites.
- Articles not relevant to the research questions.
- Articles not describing any phase of the game development life cycle.

*Study Selection*: This procedure involved two phases. In the first phase, an initial selection was made on the basis of the inclusion criteria and after reading the title, abstract, and conclusion of each article. In the second phase, if a particular article met the criteria, then the whole article was studied. One hundred twenty seven papers were identified after final selection, as shown in Fig. 3. Table 4 shows the results found in each data source and Appendix A contains a full list of selected publications.

Table 4 Results found in each data source

| Resource | Total results found | Initial selection | Final selection |
|---|---|---|---|
| IEEE Explorer | 349 | 145 | 94 |
| ACM | 120 | 30 | 17 |
| Elsevier | 200 | 38 | 15 |
| Taylor & Francis | 10 | 6 | 4 |
| Springer | 20 | 15 | 5 |
| John Wiley | 73 | 5 | 2 |
| Google Scholar | 15 | 12 | 11 |
| Total | 787 | 244 | 148 |

### 3.2.3 Quality Criteria

In this research, quality guidelines were defined based on a quality instrument that was used to assign a quality score to each article as a basis for data analysis and synthesis. The quality instrument consisted of four sections: a main section containing a generic checklist applicable to all studies, and three other sections specific to the type of study.

The checklist was based upon SLR guidelines (Kitchenham, 2004) and was derived from Kitchenham (2004) and Second expert report by American institute (2011). The detailed checklist is shown in Table 5. Some of the checklist items could be answered by "yes" or "no" and they also included a "partial" option. A value of 1 was assigned to "yes," 0 to "no," and 0.5 to "partial"; then the sum of the checklist values was used to assign a quality score to the study to assess document quality.

Table 5 Quality checklist data (Kitchenham, 2004)

| Quality Checklist | |
|---|---|
| **Generic** | |
| Are the aims clearly stated? | Yes/No |
| Was the study design appropriate with respect to its research aim? | Yes/No/Partial |
| Are statistical methods justified by the authors? | Yes/No |
| Are negative findings presented? | Yes/No/Partial |
| Are all research question answered? | Yes/No |
| Are the data collection methods adequately described? | Yes/No |
| **Empirical Analysis** | |
| Was population size reported? | Yes/No |
| Did the authors justify the sample size? | Yes/No |
| Is the sample representative of the population to which the results will be generalized? | Yes/No |
| **Theoretical Analysis** | |
| Does the author report personal observations? | Yes/No |
| Is there a link between data, interpretation, and conclusions? | Yes/No |
| Does the study cover all literature up to that point in time? | Yes/No |
| Is the focus of study reported? | Yes/No |
| **Case Study** | |
| Is the case study context defined? | Yes/No |
| Is the case study based on theory and linked to existing literature? | Yes/No |
| Is clear evidence established from observations to conclusions? | Yes/No/Partial |

### 3.2.4 Data Synthesis

For data synthesis the topics, research approaches and methods are classified and their classification details are listed below:

*Classification of topics in the GDSE Life Cycle:* This section includes a classification of the topics covered by each study with respect to the pre-production, production, and post-production phase issues involved. The 2012 ACM classification system was used for classification, which is the same method used by Kai and Card (2008). The proposed classification system has been adopted by many journals and conferences specifically for software engineering topics. The same classification was used here to classify the papers under study, and these were further fabricated based on studies found in the GDLC domain. Table 6 presents the selected classification schema.

Table 6 GDSE process life cycle classification of topics (Kai and Card, 2000)

| GDLC topics |
|---|
| **Pre-production phase** |
| Game process development management |
| Requirements Specification |
| Game system description language |
| Reusability |
| Game design document |
| Game prototype |
| Tools for designing |
| Risk Management |
| **Production Phase** |
| Assets creation |
| Story board production |
| Development platforms |
| Formal language definition |
| Programming |
| Game Engine |
| Implementation |
| **Post-production Phase** |
| Quality Assurance |
| Beta Testing |
| Usability Testing |
| Empirical Testing |
| Tools for testing |
| Marketing |

*Research Approaches and Methods Classification:* Research articles can be characterized based on their method and approach, as described by Glass et al. (2002). The main categories for scientific approach are descriptive (a system, tool, or method; a literature review can also be considered as descriptive studies), exploratory (performed where a problem was not clearly defined), and empirical (findings based on observation of its subjects). To evaluate new methods or techniques, three major empirical research methods are used: surveys, case studies, and experiments (Wohlin et al., 2000). Table 7 describes the three major empirical research types; Dyba and Dingsoyr (2008) also used the same type of empirical classification.

Table 7 Empirical methods

| Empirical method | Description |
|---|---|
| Survey | One or more questionnaires are filled out by a set of subjects either directly or by Internet, and results are derived from the answers. |
| Experiment | A specified task is performed in a highly controlled environment by a set of subjects. The results are the observations made by the subjects; in addition, task outcome inspection gives answers to research questions. |
| Case study | According to a methodology, an activity, project, or assignment is examined, and project measurements provide results. |

The data collected were statistically analyzed as follows:

- To address RQ1, the number of studies published per year, whether journal articles or conference publications, and the number of publications on the GDLC hosted by each digital library.
- To address RQ2, the major topics of the GDLC that were investigated in the software game domain.
- To address RQ3 and RQ4, the research approach or method used by number of studies.

From Section 3.2.4, data were tabulated and are presented in Appendix B.

### 3.3 Documenting (Step 10-12)

This step of the SLR describe conclusion, possible threats and limitations to the validity of this study. Authors believe that there is a chance that the word game was not part of the title of some studies, but that nevertheless they discussed game development. These studies may, therefore, have been excluded from the primary dataset by the search procedure. There are other threats that are also linked to a systematic literature review such as generalization and subjective evaluation (Shadish et al., 2002).

There are limitations to our results, although significant amounts of effort and time was spent to select the papers that were studied. More specifically, our search was limited to the academic databases. It is obvious from the results of RQ1 that developers prefer to submit their work on the blogs or forums. However, posts for different game forums and blogs cannot be included in a systematic literature review because they don't fulfil the quality criteria used for the selection of papers. In addition, the exclusion of less-known journals and conferences from the Web of Science and the Scopus index might have led to a different dataset. Finally, the classification scheme might have altered the results if they were classified by a scheme, such as the waterfall model, instead of the ACM classification scheme. Despite these limitations, the results of our systematic literature review will be useful to game development organizations and developers of software games.

## 4. Analysis of Results and Discussion

This section presents the results of statistical analysis of the data set discusses the findings concerning the RQs formulated in Section 3.1. The characteristics of the data set are tabulated for better understanding. To trace the categories of each mapped study, the interested reader is referred to the Appendix B. A total 148 studies were collated and analyzed as part of this review. To identify GDSE process life cycle domain specific characteristics, the findings of this review will be compared to results from similar studies done by Cai and Card (2008), Glass et al. (2002), and Dyba and Dingsoyr (2008).

**4.1 RQ1 What is the intensity of research activity on the GDSE process life cycle?**

Table 8 clearly shows that GDSE process life cycle research intensity has increased during the last few years. Fig. 4 showed an increase in GDSE process life cycle over time. The *y*-axis represents the number of publications in the form of a fraction and is calculated by taking year$_{(i)}$'s number of publications as the numerator and year$_{(0)}$'s number of publications as the denominator. From Table 8, 2007 was taken as year$_{(0)}$, and the first data point of the graph was calculated for year$_{(1)}$ i.e., 2008. Fig. 4 shows the results up to 2015. Years are given on the *x*-axis.

Table 8 Type of citation and per year research activity

| Citation type | Years | | | | | | | | | | | | | |
|---|---|---|---|---|---|---|---|---|---|---|---|---|---|---|
| | 2002 | 2003 | 2004 | 2005 | 2006 | 2007 | 2008 | 2009 | 2010 | 2011 | 2012 | 2013 | 2014 | 2015 |
| Book | 0 | 0 | 0 | 0 | 0 | 0 | 0 | 0 | 0 | 0 | 0 | 0 | 3 | 2 |
| Journal | 1 | 2 | 2 | 2 | 3 | 4 | 1 | 4 | 2 | 5 | 2 | 2 | 4 | 0 |
| Conference | 1 | 1 | 4 | 1 | 1 | 5 | 7 | 14 | 15 | 15 | 17 | 10 | 4 | 10 |
| Workshop | 0 | 0 | 0 | 0 | 0 | 0 | 0 | 1 | 1 | 0 | 0 | 1 | 0 | 1 |
| Total | 2 | 3 | 6 | 3 | 4 | 9 | 8 | 19 | 18 | 20 | 19 | 13 | 11 | 13 |

Fig. 4 illustrates that during the last few years, research activity in the GDSE process life cycle domain has continuously increased and the number of publications in the GDSE domain has increased at a polynomial growth rate since 2005. During 2013, 2014 and 2015 the drop in research activity is noted. It seems obvious that most of the work related to GDSE research activity was not published on the selected sources for this study. During 2014, most of the research activities were seen on the game development associations/groups web sites, like DIGRA association and Gamastura, or game developers personal blogs.

Moreover, Fig. 5 shows the list of countries most active in GDSE process life cycle topics research. Looking at research activity based on countries, China now dominates GDSE process life cycle research, but its research into the game domain started only in 2010. In four years, China has come to dominate this area of research. Before 2010, the United States and the United Kingdom were dominant.

Authors from North and South America have played a dominant role since 2004 and are still contributing in this area. Contributors in Europe also started research into the GDSE domain in 2007, but the Asian continent has dominated the GDSE domain since 2010. It can be visualized in Fig. 6. The most popular venue for GDSE research publication is IEEE; it seems that IEEE accounts for the main bulk of publications (approximately 63%), followed by Elsevier, Springer, and ACM.

## 4.2 RQ 2: What topics are being researched in the pre-production. production and post production phase?

This section addresses the identification of main research topics in the GDSE process life cycle domain. Table 9 clearly suggests that most research has been conducted in the production phase, followed by the pre-production phase. On the other hand, the post-production phase has not attracted much research interest. These GDSE process life cycle topics are somewhat different than in software engineering because of two factors: first, the GDSE domain has special needs and priorities, and second, it is a young domain which requires more fundamental research in the area of requirements, development, and coding tools. When the GDSE domain becomes mature, then other areas in the field, like testing and verification, will attract the interest of researchers.

Table 9 GDSE process life cycle topics

| GDLC topics | Frequency | Percentage |
|---|---|---|
| **Pre-production phase** | 58 | 39.18% |
| Management | 18 | 12.16% |
| Requirements specification | 9 | 6.08% |
| Game system description languages | 6 | 4.05% |
| Reusability | 3 | 2.02% |
| Game design documents | 11 | 6.75% |
| Game prototyping | 7 | 4.72% |
| Design tools | 3 | 2.02% |
| Risk management | 1 | 0.67% |
| **Production phase** | 66 | 45.27% |
| Asset creation | 7 | 4.72% |
| Storyboard production | 3 | 2.02% |
| Development platforms | 13 | 8.78% |
| Formal language definition | 2 | 1.35% |
| Programming | 17 | 11.48% |
| Game engine | 11 | 8.10% |
| Implementation | 13 | 8.78% |
| **Post-production phase** | 24 | 16.21% |
| Quality Assurance | 2 | 1.35% |
| Beta testing | 5 | 3.37% |
| Heuristic testing | 6 | 4.05% |
| Empirical testing | 2 | 1.35% |
| Test tools | 1 | 0.67% |
| Marketing | 8 | 5.40 % |

As mentioned earlier in Section 2, games have specific characteristics, which the conventional software development process cannot completely address. In the past years, research on GDSE process life cycle topics has become more active because, unlike other software products, games provide entertainment and user enjoyment, and developers need to give more importance to these aspects. As a result, research about the pre-production phase has increased. The implementation phase is shorter than in the traditional software implementation process because of the short time to market. This production-phase research intensity has attracted the interest of many researchers, and maximum research activity has been reported because the GDSE domain requires efficient development and coding techniques. McShaffrey (2003) also highlighted the importance of the production phase to counteract poor internal quality. There is much less research activity in the post-production phase than in the pre-production and production phases.

Fig. 7 presents the growth of each GDSE process life cycle research topic since 2000. It is apparent that in the pre-production phase, the most researched topic is management of the game development process, followed in this order by production-phase development platforms, programming, and implementation topics. In the post-production phase, the marketing area attracted the largest amount of research interest. The state of the art research is the description of actual primary studies, and, therefore, they are mapped according to the research topics they addressed (Budgen et al., 2008). Next, a short description of each GDSE topic is presented along with a full reference list. A full reference list of all the studies included is presented in Appendix A.

### 5.2.1 Pre-production Phase
#### 5.2.1.1 Management

In the pre-production phase, most of the studies categorized under this topic address management issues during the GDSE process life cycle. The overall management of the game development process combines both an engineering process and creation of artistic assets. Ramadan and Widyani [S1] compared various game development strategies from a management perspective, and most studies like [S3], [S6], [S7], and [S8] have proposed frameworks for game development. Game development guidelines can be followed to manage GDSE process life cycle. The presence of agile practices in the game development processes is also highlighted by some studies. Tschang [S4] and Petrillo et al. [S17] highlighted the issues in the game development process and their differences from traditional software development practices. Management of development-team members and their interaction is critically important in this aspect.

Some studies [S10] and [S11] have provided data analytics and empirical analysis of the game development process and issues of interdisciplinary team involvement. Best management practices in the game development process must consider certain elements such as staying on budget, timing, and producing the desired output. To assess game quality, five usability and quality criteria (functional, internally complete, balanced, fun, and accessible) can be used, but a process maturity model specific to the game development process is still needed to measure these processes for better management and high performance.

### 5.2.1.2 Requirements Specification

One of the main differences between the traditional software development process and GDSE process life cycle is the requirements phase. The game development process requires consideration of many factors such as emotion, game play, aesthetics, and immersive factors. In four studies, the authors have discussed the requirements engineering perspective to highlight its importance for the whole game-software development process. They discussed emotional factors, language ontology, elicitation, feedback, and emergence [S19], [S20], [S21], and [S22]. In particular, game developers must understand these basic non-functional requirements along with the game play requirements and incorporate them while developing games. The main challenges in requirements identification are a) communication between diverse background stakeholders, b) non-functional requirements incorporation with game play requirements, such as media and technology integration, and c) validation of non-functional requirement such as fun, which is very complex because it is totally dependent on the target audience. Callele et al. [S20] further fabricated a set of requirements based on emotional criteria, game-playing criteria (cognitive factors and mechanics), and sensory requirements (visual, auditory, and haptic). The requirements specification phase must address both the functional and non-functional requirements of game development.

### 5.2.1.3 Game System Description Language

Many description languages are currently used by developers, such as the UML model, agent-based methodologies, and soft-system methodologies. Quanyin et al. [S32] proposed the UML model for mobile games. They performed experiments and reported that it would be a good model for further development of games on the Android operating system. Shaker et al. [S33] extracted features of the Super Mario Brothers game from different levels, frequency sequences of level elements, and statistical design levels. Then, they analyzed the relationship between a player's experience and the level design parameters of platform games using feature analysis modelling. Tylor et al. [S28] proposed a soft system methodology for initial identification of game concepts in the development process. The proposed approach can be used instead of a popular description language because it provides an overview of the game. Chan and Yuen [S30] and Rodriguez et al. [31] proposed an ontology knowledge framework for digital game development and serious games modelling using the AOSE methodology. A system description language for games must be both intelligible to human beings and formal enough to support comparison and analysis of players and system behaviors. In addition, it must be production-independent, adequately describe the overall game process, and provide clear guidelines for developers.

### 5.2.1.4 Reusability

The existence of reusability of software (Capretz et al., 1992) and development platforms in game development has been reported by some researchers, but to gain its full advantages, commonality and variability analysis must be done in the pre-production phase. This category addresses reuse techniques for game development software (Ahmed and Capretz, 2011). Neto et al. [S34] performed a survey that analyzed game development software reuse techniques and their similarity to software product lines. Reuse techniques in game development could reduce cost and time and improve quality and productivity. For reuse techniques, commonality and variability analysis is very important, similar to a software product line. Szegletes and Forstner [S36] proposed a reusable framework for adaptive game development. The architecture of the proposed framework consisted of loosely coupled components for better flexibility. They tested their framework by developing educational games. The requirements of the new game must be well aligned with the reusable components of the previously developed game.

### 5.2.1.5 Game Design Document

The Game Design Document (GDD) is an important deliverable in the pre-production phase. It consists of a coherent description of the basic components, their interrelationships, directions, and a shared vocabulary for efficient development. Westera et al. [S37] addressed the issue of design complexity in serious games by proposing a design framework. Furthermore, Salazar et al. [S38] highlighted the importance of a game design document for game development and provided an analysis of many available game design documents from the literature. They also compared their findings with traditional software requirement specifications and concluded that a poor game design document can lead to poor-quality product, rework, and financial losses in the production and post-production phases. Hsu et al. [S40] pointed out the issues of level determination in games and trade-off decisions about them. They proposed an approach to solve the trade-off decision problem, which is based on a neural network technique and uses a genetic algorithm to perform design optimization. Khanal et al. [S41] presented design research for serious games for mobile platforms, and Cheng et al. [S42] provided design research for integrating GIS spatial query information into serious games. Finally, Ibrahim and Jaafar [S43] and Tang and Hanneghan [S44] worked on a game content model for game design documents. Currently, GDD suffers from formalism and incomplete representation; to address this issue, the formal development of GDD is very important. A comprehensive GDD (focused on the game's basic design and premises) results in good game quality.

### 5.2.1.6 Game Prototyping

Game prototyping in the pre-production phase helps the developer to clarify the fundamental mechanics of the final game. Game prototyping in the preproduction phases is considered important because it is used to convey game and play mechanics and also helps in evaluating a game player's experience. Reyno and Cubel [S49] proposed automatic prototyping for game development based on a model-driven approach. An automatic transformation generates the software prototype code in C++. De Silva et al. [S48] proposed community-driven game prototyping. The developer can approach the well-established community and focus on the technical stuff rather than starting from scratch. They used this approach for massive, multi-player online game development. Guo et al. [S50], Kanev and Sugiyam [S51], and Piesoto et al. [S52] proposed analysis of rapid prototyping for Pranndo's history-dependent games, 3D interactive computer games, and game development frameworks respectively. Prototypes also help to identify missing functionality, after which developers can easily incorporate quick design changes. Model-driven or rapid-prototyping approaches can be used to develop game prototypes.

### 5.2.1.7 Design Tools

Game design tools are used to help game developers create descriptions of effects and game events in detail without high-level programming skills. Cho and Lee [S56] and Segundo et al. [S57] proposed an event design tool for rapid game development and claimed that it does not require any kind of programming skill. These tools also enable reuse of existing components and reduce the total time of the game-creation process.

### 5.2.1.8 Risk Management

In the game development domain, risk management factors do not receive much discussion by researchers. Risk management is very important from a project management point of view. Identifying risk factors in the game development process is also important. In game development, the project manager is the game producer and must bring together management, technical, and aesthetic aspects to create a successful game. The study by Schmalz et al. [S58] is the only study highlighting the issue of risk management in video development projects. They identified two risk factors during the development process: failure of development strategy and absence of the fun factor. In game development, important risk factors can be the development strategy, the fun factor or extent of originality, scheduling, budgeting, and others, but very low priority has been given by game developers to formal analysis of risk factors.

### 5.2.2 Production Phase

#### 5.2.2.1 Asset Creation

Asset creation in the production phase is the foundation stage where game developers create the various assets and then use them in the game implementation phase. In the production phase, the first step is to create assets for the game. One of these assets is audio creation. Migneco et al. [S63] developed an audio-processing library for game development in Flash. It includes common audio-processing routines and sound-interaction Web games. Minovic et al. [S65] proposed an approach based on the model drive method for user interface development, and Pour et al. [S64] presented a brain computer interface technology that can control a game on a mobile device using EEG Mu rhythms. For audio processing, open-source libraries are available, especially for games. Audio and interface design are examples of game assets.

#### 5.2.2.2 Storyboard Production

Storyboard production is the most important phase of game production; it involves development of game scenarios for level solutions and incorporation of artificial intelligence planning techniques for representing the various features of games through a traditional white board or flow chart. Pizzi et al. [S59] proposed a rational approach that elaborated game-level scenario solutions using knowledge representation and also incorporated AI techniques to explore alternative solutions by direct interaction with generated storyboards. Finally, Anderson [S61] presented a classification of scripting systems for serious and entertainment games, and Cai and Chen [S62] explored scene editor software for game scenes. Their approach was based on the OGRE .Net framework and C++ technology. Various scripting editors based on different technologies are available for game developers to produce storyboards. Some of this software helps to develop and edit scenes at different game levels, and other software helps by generating game levels automatically based on a description.

#### 5.2.2.3 Development Platforms

The studies classified under this category proposed various types of platforms for game development. Development platforms provide a ready-made architecture for server-client connectivity and help developers create games quickly. Open-source development platforms are available, but developers must customize them according to the required functionality. Peres et al. [S69] used a scrum methodology for game development, especially for multiple platforms, and implemented interfaces with social networking Web sites such as Twitter and Facebook. Jieyi et al. [S70] proposed a platform for quick development of mobile 3D games. First, the platform implemented the game template in two environments such as the Nokia series 60 platform and the Symbian OS. The second part of the process involved analysis of the entire game structure and extraction of game parameters for later customization. Finally, the tool could be used for game customization. Lin et al. [S] developed intelligent multimedia mobile games from embedded platforms. The proposed communication protocol was able to control the embedded platform to achieve the game usability and amusement. Mao et al. [S78] presented a logical animation platform for game design and development, and Alers and Barakova [S81] developed a multi-agent platform for an educational children's game. Suomela et al. [S77] highlighted the important aspects of multi-user application platforms used for rapid game development. Some researchers have proposed a development platform similar to that described above that provides connectivity along with client customization and unnecessary updating of game servers.

### 5.2.2.4 Formal Language Description

Game semantics can be classified under formal language description for programming languages; only two studies were reported under this classification. The formal language description of game semantics provided a way to gain insight into the design of programming languages for game development. Mellies [S99] proposed a denotational prepositional linear logic for asynchronous games, and Calderon and McCusker [S100] presented their analysis of game semantics using coherence spaces. Very little work has been reported in this area, and very few game semantic descriptions of languages have been published.

### 5.2.2.5 Programming

Code complexity is increasing, especially in game development, because of the incorporation of complex modules, AI techniques, and a variety of behaviors. The most common programming languages used in game development are object-oriented structured languages such as Java, C, and C++. Studies classified under this category explored the programming aspect of game development. El Rhalibi et al. [S82] proposed a development environment based on Java Web Start and JXTA P2P technologies called Homura and NetHomura. It extends the JME game engine by facilitating content libraries, providing a new interface, and also providing a software suite that supports advanced graphical functionalities within IDE. The other two studies, done by Meng et al. [S84] and Chen and Xu [S85], also explored programming languages such as C++, DirectX, and Web GL and also Web Socket technologies for game development. Three studies by Yang et al. [S87], Yang and Zhang [S88], and Wang and Lu [S89] explored collision detection algorithms from a game logic aspect for software games, proposed A* search, and AI optimization-based algorithms.

Wang et al. [S83] proposed a framework for developing games based on J2ME technology. Zhang et al. [S92] also explored the effects of object-oriented technology on performance, executable file size, and optimization techniques for mobile games and suggested that object-oriented technology should be used with great care because the structured programming in game development is highly competitive. Bartish and Thevathayan [S86] and Fahy and Krewer [S90] analyzed the use of agents, finite state machines, and open-source libraries for the overwhelmingly complex process of multi-platform game development. Optimization techniques can be used with object-oriented programming to avoid unnecessarily redundant classes and inheritance, and to handle performance bottlenecks. These languages can be used across different development environments such as Android, iOS, Windows, and Linux. Researchers have proposed various approaches and tools for efficient game development. The integration of various development artefacts into games can also be done by generative programming, which also helps to achieve efficient development.

### 5.2.2.6 Game Engine

A game engine is a kind of special software framework that is used in the production phase for creating and developing games. Game engines consist mainly of a combination of core functionalities such as sound, a physics engine or collision detection, AI, scripting, animation, networking, memory management, and scene graphs. Hudlicka [S108] identified a set of requirements for a game engine, including identification of the player's emotions and the social interactions among game characters. This is the only study that has highlighted the important functionalities that an affective game engine must support. Another study by Wu et al. [S109] focused on game script engine development based on J2ME. It divided script engines into two types. The first type is the high-level script engine that includes packaging and refining of the script engine. The second type, the low-level script engine includes feature packages associated only with API. Four studies [S102], [S105], [S106], and [S107] explored the development of game engines on mobile platforms. Finally, Anderson et al. [S109] proposed a game engine selection tool. Recently, developers have been using previously developed or open-source game engines to economize on the game development process. Various researchers have proposed script-based, design pattern-based, and customizable game engines. In the GDSE process life cycle, game engines automate the game creation process and help a developer to develop a game in a shorter time.

### 5.2.2.7 Implementation

The foundations of game theory are used in game development because it is a branch of decision theory that describes interdependent decisions. Most studies in this category described different aspects of game implementation technologies on various types of platforms. They considered improving programming skills, 2D/3D animations and graphics, sound engineering, project management, logic design, story-writing interface design, and AI techniques. Various kinds of game implementation technologies can be found in the literature. Vanhatupa [S117] presented a survey of implementation technologies especially for browser games. The technologies explored in these studies are mainly server applications (application runtime, server-side scripting, and user interface and communication), client applications, databases, and architecture. The same study also described the accessories that can be used for implementation: application platforms, game engines, and various types of plug-ins. Abd El-Sattar [S112] proposed an interactive computer-based game framework for the implementation process. The framework includes steps from design through implementation that are based on game theory foundations and focus mainly on game models, Nash equilibrium, and strategies of play. The proposed framework includes architectural design and specifications, a proposed game overview, a game start-up interface and difficulty scaling, game modelling, the game environment and player control, and a free-style combat system.

Four studies [S113], [S114], [S119] and [S120] focused mainly on a development framework for mobile devices. Su et al. [S96] proposed a framework describing implementation of various main modules such as pressure movement, a thread pool based on the I/O completion port, and a message module. They also claimed that their proposed framework addressed the problems of traditional frameworks such as the single-server exhaustion problem, synchronization, and thread-pooling issues. Jhingut et al. [S114] discussed 3D mobile game implementation technologies from both single-player and multi-player perspectives. They also evaluated two game APIs: MDP 2.0 and M3G API. Finally, Kao et al. [S120] proposed a client framework for mobile devices that used a message-based communication protocol and reserved platform-specific data as much as possible. A few researchers have proposed agent-based frameworks as explored above for effective communication and synchronization between system components.

### 5.2.3 Post-Production Phase

#### 5.2.3.1 Quality Assurance

Process validation plays an important role in assessing game quality. Collection and evaluation of process data from the pre-production phase through to the post-production phase either provide evidence that the overall development process produces a good-quality game as a final product or reveal that it cannot. Only two studies were reported under this classification. Stacey et al. [S122] used a story-telling strategy to assess the game development process. They carried out a two-year case study on a four-person development team. Astrachan et al. [S126] tried to validate the game creation process by analyzing the development process and design decisions made during development. The scope of studies done under this category was limited. The case studies were done for small teams and were limited to only one phase. In the game development process, quality assurance and process validation are critical components, and standard methodologies are lacking. More exploration is needed to provide deeper insights. QA for games needs more research attention because very little work has been reported.

### 5.2.3.2 Beta Testing

Beta testing in games is used to evaluate overall game functionality using external testers. Beta testing is a kind of first public release for testing purposes by users. Game publishers often find it effective because bugs are identified by users that were missed by developers. If any desired functionality is missing, it must be addressed at this stage. This testing is performed before final game release. Under this classification, only four studies [S127], [S128], [S129], and [S130] were reported. Hable and Platzer [S129] evaluated their proposed development framework for mobile game platforms. Omar et al. [S128] evaluated educational computer games and identified two evaluation techniques: Playability Heuristic for Educational Games (PHEG) for expert evaluators, and Playability Assessment of Educational Games (PAEG) for real-world users. The proposed AHP-based Holistic Online Evaluation System for Educational Computer Games (AHP_HeGES) online evaluation tool can be used in the evaluation process. Very little work was reported in this category.

### 5.2.3.4 Heuristic-Based Testing

Heuristics are a kind of design guideline and can be used as an evaluation tool by game design developers or users. Basically, heuristics can be used in software engineering to test the interface. In games, evaluation must extend beyond the interface because other playability experiences also need evaluation such as the game story, play, and mechanics. Six studies [S132], [S133], [S134], [S146], [S147], and [S148] fell under this classification. Al-Azawi et al. [S132] proposed a heuristic testing-based framework for game development. The proposed framework divides testing by two types of user: experts and real-world users. Experts evaluate playability, game usability, and game quality factors. Users evaluate the game as a positive or negative experience. Omar and Jaafar [S133] and Al-Azawi et al. [S134] proposed a framework for the evaluation phase in the game development process. Heuristic testing can be done during the development process and repeated from the early design phase. It is perfect for game testing because after the game is implemented, if anything goes wrong, it will be too expensive to fix and will affect the project schedule. This topic also needs attention by researchers.

### 5.2.3.5 Empirical Testing

Empirical testing approaches for the game-testing phase have been explored by only a few researchers. The approaches described by these researchers have focused only on final-product quality and usability. Only two studies were reported under this classification [S135] and [S136]. Escudeiro and Escudeiro [S135] used a Quantitative Evaluation Framework (QEF) to evaluate serious mobile games and reported that QEF frameworks are very important in validating educational games and final-product quality. Choi [S136] analyzed the effectiveness of usability-expert evaluation and testing for game development. Experimental results showed the importance of the validation process in game development. The scope of the studies done under this category was very limited, and other aspects of final-product testing have not been explored by researchers.

### 5.2.3.6 Testing Tools

Development of testing tools has not been addressed by many researchers. Only one study [S137] was reported under this classification. Cho et al. [S137] proposed testing tools for black-box and scenario-based testing. They used their tool on several online games to verify its effectiveness. Tools for game testing facilitate the testing process. The proposed scope of study was also limited, and available testing tools have focused only on evaluation of online games.

### 5.2.3.7 Marketing

After a game has been developed, the final step is marketing. Marketing of games includes a marketing strategy and a marketing plan. The marketing strategy is directly related to the choice of users and the types of games that are in demand. The marketing plan is something that a publisher can give to a distributor to execute on the publisher's behalf. Some studies have been done from the perspective of game-user satisfaction that provide the baseline for the factors that game developers must take into account for new game development. Yee et al. [S142] described a game motivation scale based on a three-factor model that can be used to assess game trends. Three studies [S139], [S143], and [S144] empirically investigated the perspective of game-user satisfaction and loyalty. No study in the literature has directly captured a marketing strategy and a marketing plan for games.

### 4.3 RQ 3: What research approaches are being used by researchers in digital game domain?

Table 10 shows that most GDSE process life cycle studies have used an exploratory research approach. Fig. 8 shows a comparison between the three research approaches used in the GDSE process life cycle domain. Fig. 9 shows a comparison among the empirical research methods used in the GDSE process life cycle domain. The results suggest that surveys are most frequently used in GDSE domain research.

Table 10 GDLC research approach

| Research approach | Frequency |
|---|---|
| Descriptive | 61 |
| Empirical | 30 |
| Exploratory | 57 |

These results were to be expected because the GDSE domain has only been growing since 2005; before 2010 more studies follow the descriptive approach because the field was young. After 2010, more studies have followed the exploratory approach because the domain has been maturing. More specifically, exploratory and descriptive approaches seem now to be equally used in the GDSE process life cycle domain.

### 4.4 RQ4: What empirical research methods are being used in the software games domain?

Table 11 depicts the results of the RQ4. The experimental empirical method is less used in the GDSE process life cycle domain, as mentioned by Wholin et al. (2000), because carrying out formal experiments requires significant experience. The case-study method has also been used infrequently by researchers. The reason for this could be that case studies require project data obtained through various types of observations or measurements, and no research database or repository is available for the GDSE process life cycle domain. Finally, the survey method was more common than the other two methods. This is reasonable because the GDSE domain is still immature and researchers are trying to produce knowledge by questioning game users, experts, and others.

Table 11 Software games empirical research methods

| Empirical method | Frequency |
|---|---|
| Case study | 10/30 |
| Experiment | 6/30 |
| Survey | 14/30 |

## 7. Final Remarks

The GDSE process proved to be incredibly challenging as game technology including game platforms and engines changes rapidly and coding modules are used very rarely in the another game project. However, recent success of digital game industry enforces further stress along with game development challenges and highlights the need of good practices adoption for game development process. In order to find out the specific area in game development software engineering process for improvement, assessment of process activities needs to be performed. However, due to relatively young history and empirical nature of the field, there has not been any development strategies or set of best practices to carry out game development fully explored. This systematic literature review helps to identify the research gaps in game development life cycle.

The main objective of this research was to provide an insight into the GDSE process life cycle domain because, in the past, researchers have pointed out that it is different from the traditional software development process. To achieve this objective, a systematic literature review was performed, which confirmed the first step of the evidence-based paradigm. The results also confirmed that the GDSE process life cycle domain is different from the traditional software engineering development process and that research activity is growing day by day, attracting the interest of more researchers. This paper describes the various topics in the GDSE domain and highlights the main research activities related to the GDSE process life cycle. The research topics identified in the GDSE were a combination of different disciplines and together they complete the game development process.

The most heavily researched topics were from the production phase, followed by the pre-production phase. On the other hand, in the post-production phase, less research activity was reported. In the pre-production phase, the management topic accounted for the most publications, whereas in the production phase, the development platform, programming, and the implementation phase attracted the most researchers. The production phase has attracted more research because game developers focus more on implementation and programming because of the limited game-development time period. The post-production phase includes process validation, testing, and marketing topics. Very little research activity was observed in this area because the quality aspect of game development is not yet a mature field.

In addition to research topics, more researchers used exploratory research methods; as for empirical research methods, surveys were carried out by more researchers than case studies and experiments. Overall, the findings of this study are important for the development of good-quality digital games. Rapid and continual changes in technology and intense competition not only affect the business, but also have a great impact on development activities. To deal with this strong competition and high pressure, game development organizations must continually assess their activities and adopt an appropriate evaluation methodology. Use of a proper assessment methodology will help the organization identify its strengths and weaknesses and provide guidance for improvement. However, the fragmented nature of the GDSE process requires a comprehensive evaluation strategy, which has not yet been entirely explored. Finally, this kind of research work provides a baseline for other studies in the GDSE process life cycle domain and highlights research topics that need more attention in this area. The findings of this study will help researchers to identify research gaps in GDSE process life cycle and highlights areas for further research contributions. This study also is a part of a larger project aiming to propose a digital game maturity assessment model. The identified important dimensions are developer's perspective, the consumer, the business (Aleem et al. 2016), and the process itself. It also reinforces the assertion that the GDSE process life cycle domain is a complex scientific domain comparable to the software engineering development process, and it needs more attention and consideration of different factors in game development software engineering process.

**Abbreviations**

GDSE: Game Development Software Engineering (GDSE); SLR: Systematic Literature Review; GDD: Game Design Document; QEF: Quantitative Evaluation Framework

**Competing Interests**

The authors declare that they have no competing interests.

**Author's Contributions**

SA designed the study and performed the review methodology, collected the data, analyzed the data and drafted the manuscript. LC helped to conceive the study and provided guidance to carry out the quality assessments of paper, reviewed the drafted manuscript and fine-tune the final draft. FA helped in study design, provided guidance to present the analysis and helped to draft the manuscript. All authors read and approved the final manuscript.

**Figures list**

Fig. 1. SLR Steps

Fig. 2. Selection of relevant studies

Fig. 3. Study selection process

Fig. 4. Increase in GDSE process life cycle research activity

Fig. 5. Research activity per country

Fig. 6. Research activity by continent

Fig. 7. GDSE process life cycle research topics.

Fig.8 GDSE process life cycle research approaches.

Fig.9. Empirical research approaches.

# Appendix A

[S1] Ramadan, R., Widyani, Y., 2013. Game development life cycle guidelines. In *Proceedings of 5th International Conference on Advanced Computer Science and Information Systems (ICACIS)*. IEEE Computer Society, Jakarta, Indonesia, (September 28–29, 2013) 95–100.

[S2] Petrillo F., Pimenta, M., 2010. Is agility out there? Agile practices in game development. In *Proceedings of the 28th ACM International Conference on Design of Communication (SIGDOC 2010)*. ACM Digital Library, Brazil, (September 2010) 9-15.

## Appendix B Study Dataset

| Paper | Venue | Year | Country | Type of Citation | Method | Approach |
|---|---|---|---|---|---|---|
| *Pre-production* | | | | | | |
| *Game process development management* | | | | | | |
| S1 | IEEE | 2013 | Bandung | Conference | | Exploratory |
| S2 | ACM | 2010 | Brazil | Conference | | Exploratory |
| S3 | IEEE | 2006 | Korea | Conference | Survey | Empirical |
| S4 | Google Scholar | 2003 | Singapore | Conference | | Descriptive |
| S5 | Elsevier | 2004 | USA | Journal | Case study | Empirical |
| S6 | IEEE | 2010 | Malaysia | Conference | | Descriptive |
| S7 | ACM | 2011 | Portugal | Conference | | Descriptive |
| S8 | Springer | 2007 | South Africa | Journal | | Descriptive |
| S9 | IEEE | 2011 | UK | Journal | | Descriptive |
| S10 | ACM | 2012 | USA | Conference | Survey | Empirical |
| S11 | ACM | 2011 | USA | Conference | Case study | Empirical |
| S12 | IEEE | 2012 | Greece | Conference | | Exploratory |
| S13 | IEEE | 2012 | Brazil | Conference | | Exploratory |
| S14 | IEEE | 2009 | Latvia | Conference | Survey | Empirical |
| S15 | Elsevier | 2006 | Sweden | Journal | | Descriptive |
| S16 | IEEE | 2007 | USA | Conference | Survey | Empirical |
| S17 | ACM | 2009 | Brazil | Journal | Survey | Empirical |
| S18 | Tylor & Francis | 2015 | UK | Book | | Descriptive |
| *Game requirement specification* | | | | | | |
| S19 | IEEE | 2005 | Canada | Conference | | Exploratory |
| S20 | IEEE | 2010 | Canada | Workshop | Case study | Empirical |
| S21 | ACM | 2013 | USA | Workshop | Survey | Empirical |
| S22 | Tylor & Francis | 2003 | Sweden | Journal | | Descriptive |
| S23 | IEEE | 2011 | Brazil | Journal | | Exploratory |
| S24 | Elsevier | 2014 | Spain | Journal | | Exploratory |
| S25 | Springer | 2014 | Finland | Journal | | Exploratory |
| S26 | ACM | 2015 | USA | Workshop | | Descriptive |
| S27 | Google Scholar | 2015 | Korea | Conference | | Exploratory |
| *Game system description languages* | | | | | | |
| S28 | John Wiley | 2007 | UK | Journal | | Exploratory |

| S29 | Elsevier | 2013 | USA | Conference | | Descriptive |
|---|---|---|---|---|---|---|
| S30 | IEEE | 2008 | Hong Kong | Conference | Survey | Empirical |
| S31 | IEEE | 2011 | Spain | Conference | Case study | Empirical |
| S32 | Elsevier | 2011 | China | Journal | Experiments | Empirical |
| S33 | IEEE | 2011 | Denmark | Conference | | Exploratory |
| *Reusability* | | | | | | |
| S34 | IEEE | 2009 | Canada | Conference | | Descriptive |
| S35 | IEEE | 2011 | Portugal | Journal | | Descriptive |
| S36 | IEEE | 2013 | Hungary | Conference | Case Study | Empirical |
| *Game Design Document* | | | | | | |
| S37 | John Wiley | 2008 | Netherlands | Journal | | Exploratory |
| S38 | IEEE | 2012 | USA | Conference | | Descriptive |
| S39 | Tylor & Francis | 2003 | Australia | Journal | | Exploratory |
| S40 | Elsevier | 2006 | Taiwan | Journal | | Exploratory |
| S41 | IEEE | 2013 | USA | Conference | | Descriptive |
| S42 | IEEE | 2010 | China | Conference | | Descriptive |
| S43 | IEEE | 2009 | Malaysia | Conference | | Exploratory |
| S44 | IEEE | 2011 | UAE | Conference | | Descriptive |
| S45 | Tylor & Francis | 2014 | New York | Book | | Descriptive |
| S46 | Google Scholar | 2014 | Germany | Conference | | Descriptive |
| S47 | Google Scholar | 2015 | Germany | Conference | | Descriptive |
| *Prototype* | | | | | | |
| S48 | IEEE | 2013 | Japan | Journal | | Descriptive |
| S49 | ACM | 2009 | USA | Journal | | Exploratory |
| S50 | Elsevier | 2011 | China | Journal | | Descriptive |
| S51 | Elsevier | 1998 | Japan | Journal | Experiment | Empirical |
| S52 | IEEE | 2012 | Brazil | Conference | | Exploratory |
| S53 | Elsevier | 2005 | UK | Journal | Experiment | Empirical |
| S54 | IEEE | 2009 | Denmark | Journal | | Exploratory |
| *Design tools* | | | | | | |
| S55 | ACM | 2008 | Sweden | Conference | | Exploratory |
| S56 | IEEE | 2011 | Korea | Conference | | Descriptive |
| S57 | IEEE | 2010 | Brazil | Conference | | Descriptive |
| *Risk Management* | | | | | | |
| S58 | IEEE | 2014 | USA | Conference | Survey | Empirical |
| *Production* | | | | | | |
| *Storyboard production* | | | | | | |
| S59 | IEEE | 2010 | UK | Journal | | Exploratory |
| S60 | IEEE | 2013 | Finland | Conference | | Descriptive |
| S61 | IEEE | 2011 | UK | Conference | | Exploratory |
| S62 | IEEE | 2010 | China | Conference | | Exploratory |
| *Asset creation* | | | | | | |
| S63 | IEEE | 2009 | USA | Conference | | Descriptive |
| S64 | IEEE | 2008 | Australia | Conference | Survey | Empirical |
| S65 | IEEE | 2009 | Serbia | Conference | | Exploratory |
| S66 | Google Scholar | 2014 | Slovenia | Conference | | Descriptive |
| S67 | ACM | 2014 | Korea | Conference | | Exploratory |
| S68 | IEEE | 2015 | Canada | Conference | | Descriptive |
| *Development platforms* | | | | | | |
| S69 | IEEE | 2011 | Brazil | Conference | | Descriptive |
| S70 | IEEE | 2008 | China | Conference | | Exploratory |
| S71 | IEEE | 2007 | Brazil | Journal | Case study | Empirical |
| S72 | IEEE | 2009 | Brazil | Symposium | | Descriptive |

| S73 | IEEE | 2012 | Taiwan | Conference | | Descriptive |
| S74 | IEEE | 2012 | Taiwan | Conference | | Descriptive |
| S75 | IEEE | 2010 | UK | Conference | | Exploratory |
| S76 | ACM | 2004 | UK | Conference | | Descriptive |
| S77 | Springer | 2004 | Finland | Journal | | Exploratory |
| S78 | IEEE | 2010 | China | Conference | | Descriptive |
| S79 | IEEE | 2013 | Italy | Conference | | Exploratory |
| S80 | IEEE | 2010 | Spain | Conference | | Exploratory |
| S81 | IEEE | 2009 | Netherlands | Conference | Case study | Empirical |
| *Programming* | | | | | | |
| S82 | IEEE | 2009 | UK | Conference | | Exploratory |
| S83 | IEEE | 2010 | China | Conference | | Exploratory |
| S84 | IEEE | 2004 | Singapore | Conference | | Exploratory |
| S85 | IEEE | 2011 | China | Conference | | Exploratory |
| S86 | ACM | 2007 | Australia | Conference | | Exploratory |
| S87 | IEEE | 2010 | China | Conference | | Descriptive |
| S88 | IEEE | 2012 | China | Conference | | Descriptive |
| S89 | IEEE | 2010 | China | Conference | | Descriptive |
| S90 | IEEE | 2012 | Ireland | Conference | | Exploratory |
| S91 | IEEE | 2004 | Korea | Conference | | Exploratory |
| S92 | IEEE | 2007 | China | Conference | Experiments | Empirical |
| S93 | IEEE | 2012 | China | Conference | | Exploratory |
| S94 | IEEE | 2009 | Brazil | Conference | | Descriptive |
| S95 | Google Scholar | 2014 | New York | Book | | Descriptive |
| S96 | IEEE | 2014 | Malta | Journal | | Descriptive |
| S97 | Google Scholar | 2015 | New Zealand | Conference | | Exploratory |
| S98 | IEEE | 2015 | Taiwan | Conference | | Exploratory |
| *Formal language description* | | | | | | |
| S99 | Springer | 2005 | France | Journal | | Descriptive |
| S100 | Elsevier | 2010 | UK | Journal | | Descriptive |
| *Game Engine* | | | | | | |
| S101 | IEEE | 2011 | China | Conference | | Descriptive |
| S102 | IEEE | 2009 | Italy | Conference | | Descriptive |
| S103 | IEEE | 2013 | Thailand | Conference | Case study | Empirical |
| S104 | IEEE | 2012 | China | Conference | | Descriptive |
| S105 | IEEE | 2011 | China | Conference | | Exploratory |
| S106 | IEEE | 2011 | China | Conference | | Exploratory |
| S107 | IEEE | 2011 | Turkey | Conference | | Descriptive |
| S108 | ACM | 2009 | US | Conference | | Exploratory |
| S109 | IEEE | 2013 | UK | Conference | | Exploratory |
| S110 | Google Scholar | 2014 | UK | Book | | Descriptive |
| S111 | Google Scholar | 2015 | Greece | Conference | | Exploratory |
| *Implementation* | | | | | | |
| S112 | IEEE | 2008 | Egypt | Conference | | Descriptive |
| S113 | IEEE | 2009 | China | Workshop | Experiment | Empirical |
| S114 | IEEE | 2010 | Mauritius | Conference | | Exploratory |
| S115 | IEEE | 2010 | Portugal | Conference | | Descriptive |
| S116 | IEEE | 2013 | Colombia | Journal | | Descriptive |
| S117 | IEEE | 2011 | Finland | Conference | | Exploratory |
| S118 | IEEE | 2008 | Egypt | Conference | | Descriptive |
| S119 | IEEE | 2008 | Brazil | Conference | | Exploratory |
| S120 | IEEE | 2007 | Taiwan | Conference | | Exploratory |
| S121 | IEEE | 1999 | Japan | Conference | Experiment | Empirical |
| S122 | IEEE | 2015 | USA | Conference | | Descriptive |
| S123 | IEEE | 2015 | USA | Conference | | Descriptive |
| S124 | Google Scholar | 2015 | USA | Book | | Descriptive |
| | | | *Post-production* | | | |
| | | | *Quality Assurance* | | | |
| S125 | IEEE | 2007 | UK | Conference | | Exploratory |
| S126 | IEEE | 2002 | USA | Conference | | Exploratory |
| | | | *Beta Testing* | | | |
| S127 | IEEE | 2012 | Austria | Conference | | Exploratory |
| S128 | IEEE | 2012 | China | Conference | | Exploratory |
| S129 | IEEE | 2011 | Malaysia | Conference | | Exploratory |
| S130 | Elsevier | 2012 | Netherland | Journal | | Exploratory |
| S131 | ACM | 2015 | Canada | Conference | | Exploratory |
| | | | *Heuristic Testing* | | | |
| S132 | IEEE | 2013 | Oman | Conference | | Descriptive |
| S133 | IEEE | 2010 | Malaysia | Conference | | Exploratory |
| S134 | IEEE | 2013 | Oman | Conference | | Exploratory |
| S135 | Springer | 2009 | USA | Conference | | Exploratory |
| S136 | ACM | 2004 | USA | Conference | | Descriptive |
| S137 | Google Scholar | 2015 | Germany | Conference | | Descriptive |
| | | | *Empirical Testing* | | | |
| S138 | IEEE | 2012 | Portugal | Conference | Case study | Empirical |
| S139 | IEEE | 2009 | Korea | Conference | Case study | Empirical |
| | | | *Testing tools* | | | |
| S140 | IEEE | 2010 | Korea | Conference | | Descriptive |
| | | | *Marketing* | | | |
| S141 | IEEE | 2012 | USA | Conference | | Descriptive |
| S142 | Elsevier | 2014 | Taiwan | Journal | Survey | Empirical |
| S143 | ACM | 2007 | Australia | Journal | | Exploratory |
| S144 | Elsevier | 2006 | France | Journal | | Descriptive |
| S145 | ACM | 2012 | USA | Conference | Survey | Empirical |
| S146 | Elsevier | 2012 | UK | Journal | Survey | Empirical |
| S147 | Elsevier | 2009 | Taiwan | Journal | Survey | Empirical |
| S148 | IEEE | 2009 | China | Conferences | | Descriptive |